\documentclass[reprint,amsmath,amssymb,aps,prl,longbibliography]{revtex4-2}

\usepackage{graphicx}
\usepackage{dcolumn}
\usepackage{bm}
\usepackage{xcolor}
\usepackage{footmisc}
\usepackage[colorlinks=true,allcolors=blue]{hyperref}
\usepackage{comment}

\begin{document}

\preprint{APS/123-QED}

\title{Inverse supersymmetry in finite temperature Bose-Fermi mixtures}

\author{Zachary Gazzillo}
\author{Carlos A. R. S\'a de Melo}
\affiliation{School of Physics, Georgia Institute of Technology, Atlanta, Georgia 30332, USA.
}
\date{\today}

\begin{abstract}
    We investigate nearly degenerate Bose-Fermi mixtures and show that the breaking of generalized supersymmetry (gSUSY) between bosons and fermions, with up to two internal states, manifests itself through the emergence of fermionic Goldstino modes with up to four flavors. In particular, we draw a distinction between typical supersymmetry (SUSY), where bosons have pseudospin 0 and fermions have pseudospin 1/2, and inverse supersymmetry (iSUSY), where bosons have pseudospin 1/2 and fermions have pseudospin 0. In such systems, we highlight that the Goldstino pseudospin is carried by either its constituent fermion (SUSY) or boson (iSUSY). We then distinguish between these two cases by depicting their differing effects on the spectral functions of the bosonic and fermionic atomic species. Lastly, we propose radio-frequency- or microwave-spectroscopy experiments, analogous to momentum (angular) resolved photoemission in condensed matter physics, to measure the pseudospin-dependent spectral functions and detect the emergence of Goldstino modes in mixtures of $^{39}$K and $^{40}$K.

\end{abstract}

\maketitle


\textit{Introduction\textemdash}Supersymmetry (SUSY), that symmetry which relates to the exchange of bosonic and fermionic particles, has emerged as one of the leading beyond-standard-model (BSM) theories since its effective introduction by Wess and Zumino's construction of their eponymous, linearly realized model \cite{Wess_1974}. These SUSY theories are motivated by their potential to address several open problems in fundamental physics, including the hierarchy problem (related to the mass of the Higgs boson) and the nature of dark matter, though no experimental evidence has yet been produced \cite{Constantin_2025}. Due to SUSY's apparent ability to uncover a rich phenomenology, the extension of non-SUSY models into a supersymmetric regime has become a natural avenue of exploration for systems containing both bosonic and fermionic degrees of freedom \cite{snegirev_2025,Behrends_2020}.

Resulting from the growing ability to prepare atomic mixtures of ultracold quantum gases with tunable interactions in a single trap, there has been an explosion of interest in mixtures of ultracold bosonic and fermionic atomic gases~\cite{Smith_2000,Viverit_2002,Fehrmann_2004,Illuminati_2004,Batrouni_2007,Modugno_2007,sademelo_2009,Pieri_2013,Mathy_2013,Grimm_2018,Bruun_2018,Ozeri_2020,Mueller_2021,Wu_2024,Pieri_2025,knap_2025}.
In addition to recent successes in experimentally realizing stable Bose-Fermi mixtures \cite{Ferrier-Barbut_2014, Ikemachi_2016, Laurent_2017, Wu_2011, Bochenski_2024,  van_Abeelen_1997, Schreck_2001, Truscott_2001, Hadzibabic_2002, Roati_2002, Zaccanti_2006, DeSalvo_2019,Luo_2023,Lippi_2024,Zwierlein_2024}, such mixtures also provide a system that can be intuitively extended into the supersymmetric regime \cite{Yu_2008, Shi_2010, Blaizot_2015, Bradlyn_2016, Blaizot_2017, Tajima_2021, Zhang_2024} thus allowing for the imagining of tabletop experiments with realizable SUSY. With this, one is able to utilize ultracold atoms as quantum simulators of the more fundamental BSM physics. Further, when such a system is near-supersymmetric, a fermionic Nambu-Goldstone-like mode emerges, referred to as a ``Goldstino'' \cite{Kratzert_2003,Lebedev_1989,Salam_1974,Witten_1981}. These near-supersymmetric mixtures can be potentially realized experimentally by combinations of stable, consecutive isotopes to minimize mass differences (e.g. $^{7}\text{Li}$-$^{6}\text{Li}$ \cite{Ferrier-Barbut_2014, Laurent_2017, Ikemachi_2016}, $^{41}\text{K}$-$^{40}\text{K}$ \cite{Wu_2011}, and $^{39}\text{K}$-$^{40}\text{K}$ 
\cite{Bochenski_2024}). 
In recent years, Goldstinos have been studied for several different parameter ranges and temperatures \cite{Yu_2008, Shi_2010, Blaizot_2015, Bradlyn_2016, Blaizot_2017, Tajima_2021, Zhang_2024}. However, in every instance, such theories rely on pseudospinless mixtures, in which both the bosons and the fermions lack spin (internal) degrees of freedom leading to a single Goldstino mode. This simplification does not directly connect to BSM theories, where the fermions in supersymmetric pairs and the Goldstino mode carry spin \cite{Sohnius_1985}.

In this paper, we outline a theory of interacting, nearly degenerate bosons and fermions with pseudospin (hereafter referred to as \textit{spin}) with up to two internal states, at temperatures above the bosons' condensation temperatures, both near and far from supersymmetry. We call this theory 
generalized supersymmetry (gSUSY), 
for which there is a maximum of four Goldstino flavors.

We discuss two particular examples of gSUSY which are experimentally friendly: (a) a SUSY case in which the mixture contains \textit{spin}-0 bosons and \textit{spin}-1/2 fermions such that the \textit{spin} of the resulting Goldstinos is carried by the component fermion, and (b) an inverse SUSY (iSUSY) case in which the mixture contains \textit{spin}-1/2 bosons and \textit{spin}-0 fermions such that the \textit{spin} of the resulting Goldstinos is carried by the component boson. We derive and explore the emergence of these excitations in these two cases to motivate the understanding of iSUSY as a phenomenon that is unique to Bose-Fermi atomic mixtures, since fundamental bosons cannot have half-integer spins, and thus iSUSY is not allowed in BSM theories, while SUSY is.

While our analytical results apply to a general Bose-Fermi mixture where \textit{spin} plays an important role, our numerical calculations of Goldstino dispersions and fermionic  or bosonic spectral functions, measured in 
radio frequency-~\cite{Stewart_2008,Koschorreck_2012,Jin_2015} and microwave-~\cite{Li_2024} spectroscopy experiments, are compatible with parameters reflective of a $^{39}\text{K}$-$^{40}\text{K}$ mixture.


\textit{Hamiltonian\textemdash} We consider a non-relativistic Bose-Fermi mixture of atomic gases with up to two internal states in three dimensions. We assume a uniform particle density, which can be achieved experimentally with a box potential \cite{Navon_2021}, and use units where 
$\hbar=k_B=1$. This system is described by the Hamiltonian, 
\begin{equation}
    \label{Eq: Hamiltonian density}
    H=H_b+H_f+V,
\end{equation}
where the kinetic energy of the bosons is 
\begin{equation}
    H_b=\int d^3\bm{x}\sum_{r} b_r^\dagger(\bm{x})\left[-\frac{\nabla^2}{2m_b}-\mu_r^b\right]b_r(\bm{x})
\end{equation}
and the kinetic energy of the fermions is
\begin{equation}
    H_f=\int d^3\bm{x}\sum_{r} f_r^\dagger(\bm{x})\left[-\frac{\nabla^2}{2m_f}-\mu_r^f\right]f_r(\bm{x}),
\end{equation}
with $b_r(\bm{x})$ ($f_r(\bm{x})$) being the field operators for bosons (fermions) in internal state $r$. The particles have mass $m_{b(f)}$ and chemical potential 
$\mu^{b(f)}_r$. Further, the interaction potential for a dilute Bose-Fermi mixture is
\begin{eqnarray}
    \nonumber
    V &=& \int d^3\bm{x}\sum_{rs}\Bigg[\frac{1}{2}U^{bb}_{rs}b_r^\dagger(\bm{x})b^\dagger_s(\bm{x})b_s(\bm{x})b_r(\bm{x})\\*\nonumber
    &+& \frac{1}{2}U^{ff}_{rs}f_r^\dagger(\bm{x})f^\dagger_s(\bm{x})f_s(\bm{x})f_r(\bm{x}) \\* 
    &+& U^{bf}_{rs}b_r^\dagger(\bm{x})f^\dagger_s(\bm{x})f_s(\bm{x})b_r(\bm{x})\Bigg],
\end{eqnarray}
where $U^{\gamma_1\gamma_2}_{rs}$, with dimensions of energy times volume,  is the contact interaction strength between particles $\gamma_1$ in internal state $r$ and $\gamma_2$ in internal state $s$ ($\gamma_1,\gamma_2\in\{b,f\}$). The sum is over internal states $r,s\in\{\uparrow,\downarrow\}$ when describing a \textit{spin}-1/2 particle or $\in\{0\}$ when describing a \textit{spin}-0 particle. 

Under this model, there exist particular locations in parameter space referred to as generalized supersymmetry (gSUSY) points, where the supercharges
\begin{equation}
    Q_{ij}=\int d^3\bm{x}b_i^\dagger(\bm{x})f_j(\bm{x})
\end{equation}
are conserved. The integrand $q_{ij}(\bm{x})= b_i^\dagger(\bm{x})f_j(\bm{x})$ are local operators that exchange a boson in state $i$ for a fermion in state $j$. These operators are also the generators of excitations that break gSUSY, that is, the field operator for the Goldstinos labeled by $(i,j)$. Conservation of these supercharges requires
\begin{eqnarray}
    \nonumber
    \left[H,Q_{ij}\right]&=&\int d^3\bm{x}b_i^\dagger(\bm{x})\Bigg[ \frac{\chi\nabla^2}{2m^*}
    +\sum_{r}\Delta U_{irrj}^bb_r^\dagger(\bm{x})b_r(\bm{x})\\*
    &+&\Delta\mu_{ij}+\sum_{r}\Delta U_{irrj}^ff_r^\dagger(\bm{x})f_r(\bm{x})\Bigg]f_j(\bm{x}) \label{Eq: Supercharge Conservation}
\end{eqnarray}
to evaluate to zero, where $m^* = m_bm_f/(m_b+m_f)$ is the reduced mass. Here, the gSUSY breaking quantities are defined to be the mass-difference ratio $\chi =(m_b-m_f)/(m_b+m_f)$, the chemical potential differences $\Delta\mu_{ij}=\mu^f_j-\mu^b_i$, and the interaction strength differences $\Delta U^b_{irrj}= U^{bb}_{ir}-U^{bf}_{rj}$ and $\Delta U^f_{irrj}= U^{bf}_{ir}-U^{ff}_{rj}$. If all these quantities can be tuned to zero, then gSUSY between species in states $i$ and $j$ is achieved, or in particular SUSY and iSUSY. If at least one of these quantities is not zero, then there could be a maximum of
four Goldstino modes for gSUSY, and a maximum of two Goldstinos modes for SUSY, where fermions carry the \textit{spin}, and iSUSY, where bosons carry the \textit{spin}.


\begin{figure}
    \includegraphics[width=8 cm]{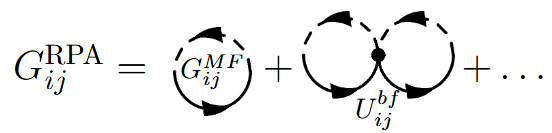}
    \caption{\label{Fig: Goldstino Diagrams} The RPA Bose-Fermi correlator   $G_{ij}^\text{RPA}$, the poles of which form the Goldstino propagator $G^q_{ij}$, consisting of the mean-field Bose-Fermi correlator $G^\text{MF}_{ij}$, and products thereof connected by the Bose-Fermi interacting strength, $U^{bf}_{ij}$. Here, the solid (dashed) lines represent the independent fermion (boson) propagator. 
    }
\end{figure}

\textit{Goldstino and Anti-Goldstino Emergence\textemdash} To investigate gSUSY, SUSY, and iSUSY breaking for a realistic Bose-Fermi mixture, we consider $m_b \ne m_f$, that is, $\chi \neq 0$ and obtain the dispersion of Goldstino modes in the presence and absence of other gSUSY, SUSY and iSUSY breaking parameters. We study a three-dimensional system at temperatures $T$ slightly larger than the condensation temperature of bosons $T_{\text{BEC},i}$ and of the order
of the Fermi temperatures $T_{\text{F},j}$. We assume that all interactions are repulsive to ensure the stability of the bosonic system, to avoid superfluidity in the fermion sector, and to favor  Goldstinos and anti-Goldstinos over Bose-Fermi molecular bound states.

We begin with the mean-field Bose-Fermi correlator $G_{ij}^\text{MF}(\bm{p},\omega)$, which is represented by the one-loop diagram in the first line of Fig. \ref{Fig: Goldstino Diagrams} describing a bubble composed of a fermion particle and a boson hole  (See End Matter). This diagram produces
\begin{equation}
G_{ij}^\text{MF}(\bm{p},\omega)=
\int \frac{d^3\bm{k}}{(2\pi)^3}
\frac{n_b\left(\xi^b_i\left(\bm{k}-\bm{p}\right)\right)+n_f\left(\xi^f_j\left(\bm{k}\right)\right)}{\omega+\xi^b_i\left(\bm{k}-\bm{p}\right)-\xi^f_j\left(\bm{k}\right)+\text{i}0^+},
\label{Eq: GMF}
\end{equation}
where $\omega$ is a frequency and $n_{b(f)}$ is the Bose-Einstein (Fermi-Dirac) distribution.  
We note that $G_{ij}^\text{MF} (\bm{p}, \omega)$ has dimensions of
inverse energy times inverse volume.
Further, $\xi^{b(f)}_r(\bm{k})=|\bm{k}|^2/(2m_{b(f)})-\mu^{b(f)}_r+\Sigma^{b(f)}_r$ is the single particle energy of a boson (fermion) with \textit{spin} $r$ and momentum $\bm{k}$ where
\begin{eqnarray}
\Sigma^b_i&=& \sum_r\left[U^{bb}_{ir}N_r^b(1+\delta_{ir})+U^{bf}_{ir}N^f_r\right]\\
\Sigma^f_j&=& \sum_r\left[U^{bf}_{rj}N^b_r+U^{ff}_{rj}N^f_r(1-\delta_{rj})\right]
\end{eqnarray}
are the Hartree self-energies and $N^{b(f)}_r$ is the particle density of bosons (fermions) in \textit{spin} state $r$.

Performing the Random Phase Approximation (RPA) shown in Fig. \ref{Fig: Goldstino Diagrams}, we obtain the RPA Bose-Fermi correlator
\begin{equation}
\label{Eq: RPA}
G^\text{RPA}_{ij}(\bm{p},\omega)=\frac{G_{ij}^\text{MF}(\bm{p},\omega)}{1-U^{bf}_{ij}G_{ij}^\text{MF}(\bm{p},\omega)}.
\end{equation}
The Goldstinos emerge as poles of  
$G^\text{RPA}_{ij}(\bm{p},\omega)$ represented by the integer zeros of the denominator, and the fermion-particle--boson-hole (or simply particle-hole) continuum is represented by a branch cut. The equation for the zeros of the denominator is shown in the End Matter to be equivalent to
\begin{equation}
\label{Eq: Pole eq}
\omega-\Omega_{ij}-\Lambda_{ij}(\bm{p},\omega)= 0,
\end{equation}
where a factor of $U^{bf}_{ij}N_{ij}$ has been divided out, in which $N_{ij}=N^b_i+N^f_j$ is the sum of the particle densities for the relevant \textit{spin} states. 
Written in this way, the pole of the RPA Bose-Fermi correlator corresponding to the $(i,j)$ Goldstino's dispersion $\omega=\mathcal{D}_{ij}(\bm{p})$ is split into an approximate, zero-momentum solution and non-zero-momentum corrections. 
The zero-momentum solution is 
$\omega = \mathcal{D}_{ij}(\bm{0})$, which, to first order in the gSUSY breaking parameters, can be approximated by 
$\omega \approx\Omega_{ij}$ as shown in the End Matter. Here and in Eq.~(\ref{Eq: Pole eq}), 
\begin{eqnarray}
\nonumber\Omega_{ij}&=&-\Delta\mu_{ij}+\chi\langle\mathcal{E}_{ij}\rangle-\sum_r\Delta U^b_{irrj}N^b_r(1+\delta_{ri})\\*&-&\sum_r\Delta U^f_{irrj}N^f_r(1-\delta_{rj}),
\label{Eq: Omega}
\end{eqnarray}
which is linear in the gSUSY breaking parameters,
$\Delta\mu_{ij}$, $\chi$, $\Delta U^b_{irrj}$, and $\Delta U^f_{irrj}$,
appearing in Eq.~(\ref{Eq: Supercharge Conservation}).
The term
\begin{equation}
\label{Eq: E def}
\langle\mathcal{E}_{ij}\rangle=\frac{m_b\langle K_b\rangle+m_f\langle K_f\rangle}{N_{ij}m^*}
\end{equation}
is a mass-averaged kinetic energy per particle, with $\langle K_{b(f)}\rangle$ being the average kinetic energy density of the bosons (fermions). 
The second term in Eq.~(\ref{Eq: Pole eq}) 
is
\begin{eqnarray}
    \nonumber
    N_{ij}\Lambda_{ij}(\bm{p},\omega)=\frac{|\bm{p}|^2}{2m_f}\left(N^b_i-N^f_j\frac{1-\chi}{1+\chi}\right)\\*
    +\int \frac{d^3\bm{k}}{(2\pi)^3}\frac{n_{ij}(\bm{k},\bm{p})\left[\omega-\zeta_{ij}(\bm{k},\bm{p})-U^{bf}_{ij}N_{ij}\right]^2}{\omega - \zeta_{ij}(\bm{k},\bm{p})+\text{i}0^+},
    \label{Eq: Lambda Def}
\end{eqnarray}
corresponding to a non-zero momentum contribution. Finally, $\zeta_{ij}(\bm{k},\bm{p}) = \Omega_{ij} - \chi\langle\mathcal{E}_{ij}\rangle-U^{bf}_{ij}N_{ij} +\frac{\chi |\bm{k}|^2}{2m^*}
-\frac{|\bm{p}|^2}{2\Delta m}$
is associated with the energy of unbound fermion particles and boson holes, $n_{ij}(\bm{k},\bm{p})=n_b\left(\xi^b_i\left(\bm{k}-\frac{m_b}{\Delta m}\bm{p}\right)\right)+n_f\left(\xi^f_j\left(\bm{k}-\frac{m_f}{\Delta m}\bm{p}\right)\right)$
represents the sum of bosonic and fermionic occupations, $\Delta m=m_b-m_f$ is the mass difference, and $\text{i}0^+$ is a small imaginary value which goes to zero. Here, we emphasize that $\Lambda_{ij}(\bm{0},\Omega_{ij})$, as it appears in the approximate zero-momentum solution 
$\mathcal{D}_{ij}(0)=\Omega_{ij}+\Lambda_{ij}(\bm{0},\mathcal{D}_{ij}(\bm{0}))$, is of $O(\chi^2)$, see End Matter.

\begin{table}
    \caption{Parameters used in the numerical calculations for a Bose-Fermi mixture of $^{39}$K (boson) and $^{40}$K (fermion). All unlisted scattering lengths are set to zero, and the length scale $a_0$ is the Bohr radius. Finally, the chemical potentials, non-interacting condensation temperatures, and independent Fermi temperatures are numerically calculated and consistent with the listed parameters.}
    \centering
    \resizebox{\linewidth}{!}{%
    \begin{tabular}{l@{\hspace{2mm}}r@{\hspace{0.5mm}}lcl@{\hspace{2mm}}r@{\hspace{0.5mm}}l}
        \hline\hline
        \multicolumn{3}{c}{\textbf{SUSY}} & \hspace{4mm} & \multicolumn{3}{c}{\textbf{iSUSY}} \\ \hline
        \textbf{Quantity} & \multicolumn{2}{c}{\textbf{Value}} & & \textbf{Quantity} & \multicolumn{2}{c}{\textbf{Value}} \\ 
        T & $1.22\times10^{-7}$ & K &  & T & $8.58\times10^{-8}$ & K \\
        $T_{\text{BEC},0}$ & $1.21\times10^{-7}$ & K & & $T_{F,0}$ & $3.53\times10^{-7}$ & K \\ 
        $T_{F,\uparrow}$ & $2.69\times10^{-7}$ & K & & $T_{\text{BEC},\uparrow}$ & $3.53\times10^{-8}$ & K \\
        $T_{F,\downarrow}$ & $1.70\times10^{-7}$ & K & & $T_{\text{BEC},\downarrow}$ & $8.57\times10^{-8}$ & K \\
        $m_b$ & 39 & amu & & $m_b$ & 39 & amu \\
        $m_f$ & 40 & amu & & $m_f$ & 40 & amu \\
        $a_{0\uparrow}^{bf}$ & 187 & $a_0$ & & $a_{\uparrow0}^{bf}$ & 106 & $a_0$ \\ 
        $a_{0\downarrow}^{bf}$ & 149 & $a_0$ & & $a_{\downarrow0}^{bf}$ & 126 & $a_0$ \\ 
        $N^b_0$ & $5\times10^{12}$ & cm$^{-3}$ & & $N^f_0$ & $7.5\times10^{12}$ & cm$^{-3}$ \\ 
        $N^f_\uparrow$ & $5\times10^{12}$ & cm$^{-3}$ & & $N^b_\uparrow$ & $3\times10^{12}$ & cm$^{-3}$ \\ 
        $N^f_\downarrow$ & $2.5\times10^{12}$ & cm$^{-3}$ & & $N^b_\downarrow$ & $2\times10^{12}$ & cm$^{-3}$ \\ 
        $\mu^b_0$ & $0.00615$ & $\epsilon_F$ & & $\mu^f_0$ & $1.53$ & $\epsilon_F$ \\ 
        $\mu^f_\uparrow$ & $1.01$ & $\epsilon_F$ & & $\mu^b_\uparrow$ & $0.0336$ & $\epsilon_F$ \\ 
        $\mu^f_\downarrow$ & $0.388$ & $\epsilon_F$ & & $\mu^b_\downarrow$ & $0.0115$ & $\epsilon_F$ \\    \hline\hline
    \end{tabular}%
    }
    \label{Tab:parameters_all}
\end{table}

In this form, the branch cut in $G^\text{RPA}_{ij}(\bm{p},\omega)$ manifests as the region in $(\bm{p},\omega)$-space in which $\Lambda_{ij}$ takes on an imaginary component, defining a semi-infinite region 
\begin{equation}
\label{Eq: Continuum definition}
\begin{cases}
\omega\leq\zeta_{ij}(\bm{0},\bm{p}), & \chi<0 \\
\omega\geq\zeta_{ij}(\bm{0},\bm{p}), & \chi>0
\end{cases}
\end{equation}
wherein the Goldstino acquires a finite lifetime and becomes unstable. The boundary of this region at zero-momentum $\zeta_{ij}(\bm{0},\bm{0})$ thus represents the threshold energy of the dissociated boson hole ($i$) and fermion particle ($j$). This implies that energetically stable Goldstinos only emerge when their dispersions $\mathcal{D}_{ij}(\bm{p})$ exist outside of the continuum at some momentum, in addition to breaking gSUSY.
For $\chi \ne 0$ and within the RPA approach described above, a necessary and sufficient condition to have a zero momentum stable Goldstino is $1<-U^{bf}_{ij}N_{ij}/(\chi\langle\mathcal{E}_{ij}\rangle)$. For $\chi<0$ ($^{39}$K-$^{40}$K), this relation describes a positive energy cost $\Delta\omega_{ij}(\bm{p}=\bm{0})=\mathcal{D}_{ij}(\bm{0})-\zeta_{ij}(\bm{0},\bm{0})$ of creating the Goldstino that is equal to the difference between the Goldstino energy $\mathcal{D}_{ij}(\bm{0})$ and 
the Bose-Fermi continuum threshold 
$\zeta_{ij}(\bm{0},\bm{0})$.

Alternatively, in mixtures where $\chi>0$ ($^{7}\text{Li}$-$^{6}\text{Li}$ and $^{41}\text{K}$-$^{40}\text{K}$), poles in the RPA correlator exist only when $\mathcal{D}_{ij}(\bm{p})$ is less than $\zeta_{ij}(\bm{0},\bm{0})$. In such cases, their energy being below the particle-hole continuum implies that these poles cannot be interpreted as Goldstinos. Instead the true excitations are anti-Goldstinos with the field operator $\tilde{q}_{ij}(\bm{x})=q^\dagger_{ij}(\bm{x})=b_i(\bm{x})f_j^\dagger(\bm{x})$, the equations for which are all identical to those of the Goldstino under a $(\bm{p},\omega)\rightarrow(-\bm{p},-\omega)$ transformation.

For brevity, we focus on stable Goldstino excitations when gSUSY is broken. Outside of the particle-hole continuum, we approximate the RPA correlator $G_{ij}^\text{RPA}$ to consist of only its pole and is therefore only the $(i, j)$ Goldstino propagator
\begin{equation}
G^q_{ij}(\bm{p},\omega) = \frac{Z_{ij}(\bm{p})}{\omega-\mathcal{D}_{ij}(\bm{p})+\text{i}0^+}
\label{Eq: Gq}
\end{equation}
with the residue of the pole for the $(i,j)$ Goldstino (proportional to the wavefunction renormalization) being
\begin{equation}
Z_{ij}(\bm{p})=\left[\frac{U^{bf}_{ij}}{G^\text{MF}_{ij} (\bm{p})}
\int\frac{d^3\bm{k}}{(2\pi)^3}\frac{n_{ij}(\bm{k},\bm{p})}{[\mathcal{D}_{ij}(\bm{p})-\zeta_{ij}(\bm{k},\bm{p})]^2}\right]^{-1}.
\label{Eq: residue}
\end{equation}
 Here, $G^\text{MF}_{ij} (\bm{p})
= G^\text{MF}_{ij} (\bm{p},\mathcal{D}_{ij}(\bm{p}))$. See the End Matter for a discussion of the derivation and limits of $Z_{ij}(\bm{p})$.
Note that there are up to four Goldstinos modes, labeled by the boson \textit{spin} index 
$i \in \{\uparrow, \downarrow \}$ and 
the fermion \textit{spin} index 
$j \in \{\uparrow, \downarrow \}$.

\begin{figure}
\includegraphics[width=8.5 cm]{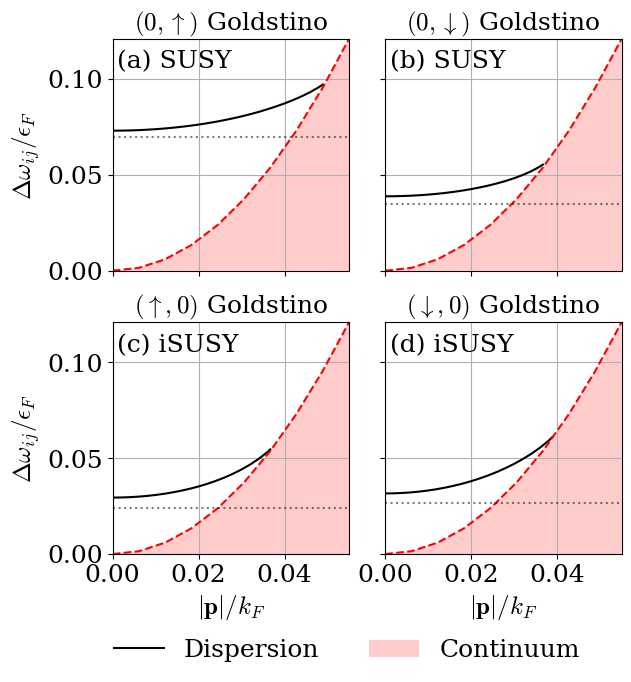}
\caption{\label{Fig: Dispersions} 
Plots of the energy difference 
$\Delta\omega_{ij} = \omega - \zeta_{ij}(\bm{0},\bm{0})$ between the energy of excitations $\omega$ and the zero-momentum particle-hole continuum boundary 
$\zeta_{ij} (\bm{0},\bm{0})$ versus momentum modulus 
$\vert \bm{p} \vert$. In each plot, $\epsilon_F$ and $k_F$ represent the Fermi energy and Fermi momentum, respectively.
We use parameters of Tab.~\ref{Tab:parameters_all} to illustrate SUSY in panels (a) and (b), as well as iSUSY in panels (c) and (d).
The solid black lines represent the Goldstino dispersions $\omega = \mathcal{D}_{ij}(\bm{p})$, while the shaded pink area describes the particle-hole continuum with the dashed red line being its boundary. Finally, the dotted gray line is the approximate zero-momentum solution 
$\omega = \mathcal{D}_{ij}(\bm{0})\approx\Omega_{ij}$.
}
\end{figure}
%


\textit{Numerical Results\textemdash} So far, we have discussed the general case of a weakly interacting, nearly degenerate Bose-Fermi mixture, where gSUSY can be broken in many ways. Moving forward, we focus on 
the simplest nontrivial examples that include \textit{spins}, that is, SUSY (\textit{spin} is carried by fermions) and iSUSY \textit{spin} is carried by bosons). 
Further, we focus on the exemplary case in which the interspecies interaction is much stronger than the intraspecies interactions; as such, we make the equivalent approximation and let $U^{bb}_{r s}=U^{ff}_{rs}=0$.
Thus, here, we replace $U^{bf}_{rs}$ by it corresponding
scattering length $a^{bf}_{rs}$, see End Matter.

So far, it has been experimentally difficult to study in detail Bose-Fermi mixtures of different isotopes from the same element~\cite{Ferrier-Barbut_2014, Laurent_2017, Ikemachi_2016, Wu_2011}. However,
more recently some progress has been made for $^{39}\text{K}$-$^{40}\text{K}$ mixtures~\cite{Bochenski_2024}, thus we use the parameters described in Table \ref{Tab:parameters_all} to illustrate the simplest example of SUSY and iSUSY including \textit{spins} where Bose-Fermi interactions are dominant. 
We define momentum and energy scales via the total density of fermions $N^f_T = \sum_j N^f_j$ to preserve their value across SUSY and iSUSY examples, in which case we write our ``Fermi'' momentum as $k_F=(3\pi^2N^f_T )^{1/3}$, and our ``Fermi'' energy as $\epsilon_F=k_F^2/(2m_f)$.

In Fig. \ref{Fig: Dispersions}, we plot 
the energy difference 
$\Delta\omega_{ij} = \omega - \zeta_{ij}(\bm{0},\bm{0})$ between the energy of excitations $\omega$ and the zero-momentum particle-hole continuum boundary 
$\zeta_{ij} (\bm{0},\bm{0})$ versus momentum modulus 
$\vert \bm{p} \vert$. We use parameters of Tab.~\ref{Tab:parameters_all} to illustrate SUSY in panels (a) and (b), as well as iSUSY in panels (c) and (d).
The solid black lines are the Goldstino dispersions $\omega = \mathcal{D}_{ij}(\bm{p})$, while the shaded pink area represents the particle-hole continuum with the dashed red line being its boundary. Finally, the dotted gray line demarks the approximate zero-momentum solution 
$\omega = \mathcal{D}_{ij}(\bm{0})\approx\Omega_{ij}$.
Panels (a) and (b) show the 
\textit{spin}-up $(0, \uparrow)$ and \textit{spin}-down $(0, \downarrow)$ Goldstino dispersions for the SUSY case, where fermions carry the \textit{spin}.
Panels (c) and (d) show the 
\textit{spin}-up $(\uparrow, 0)$ and \textit{spin}-down $(\downarrow, 0)$ Goldstino dispersions
for the iSUSY case, where the bosons carry the
\textit{spin}. 

\begin{figure}
    \includegraphics[width=8.5 cm]{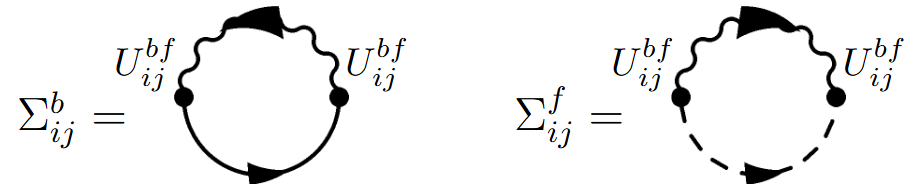}
    \caption{\label{Fig: Self Energy Diagram} The self-energy diagrams $\Sigma^{b(f)}_{ij}$ contributing to the energy of a boson (fermion) in \textit{spin} $i$ $(j)$ emerging due to the existence of the $(i,j)$ Goldstino with propagator $G^q_{ij}$ represented by the wavy line. The solid (dashed) lines describe the mean-field corrected fermion (boson) propagator.}
\end{figure}
%


\textit{Boson Spectral Functions\textemdash}The most immediate impact of the Goldstinos is seen in the spectral function of the constituent particles, which can be measured using radio-frequency (RF)~\cite{Stewart_2008,Koschorreck_2012,Jin_2015} or microwave~\cite{Li_2024} photons that eject atoms in analogy with photoemission experiments in condensed matter. Here, we only discuss the bosonic spectral function, deferring the fermionic version to the End Matter. 

The spectral function for bosons with \textit{spin} $i$ is
$\mathcal{A}^{b}_{i}(\bm{p},\omega)
=-\Im \left[ G_{i}^{b}(\bm{p},\omega)\right]/\pi$, where 
$\Im$ is the imaginary part of the bosonic retarded propagator 
$G^{b}_{i}(\bm{p},\omega)=[\omega-\xi^{b}_{i}(\bm{p})-\sum_r\Sigma^b_{ir}(\bm{p},\omega)+\text{i}0^+]^{-1}$.
This includes the self-energy corrections originating from the fermion-Goldstino interactions displayed in Fig. \ref{Fig: Self Energy Diagram}, which take the form
\begin{equation}
    \Sigma^b_{ir}(\bm{p},\omega)\approx\left(U^{bf}_{ir}\right)^2\int \frac{d^3\bm{k}}{(2\pi)^3}Z_{ir}(\bm{k})\Pi_b(\bm{k},\bm{p},\omega),
\end{equation}
where the fermion-Goldstino term 
\begin{equation}
    \Pi_b(\bm{k},\bm{p},\omega)=\frac{n_f(\xi^f_r(\bm{p}+\bm{k}))-n_f(\mathcal{D}_{ir}(\bm{k}))}{\omega-\xi^f_r(\bm{p}+\bm{k})+\mathcal{D}_{ir}(\bm{k})+i0^+}
\end{equation}
describes the boson's self-energy contributions. 

\begin{figure}
\includegraphics[width=8.5 cm]{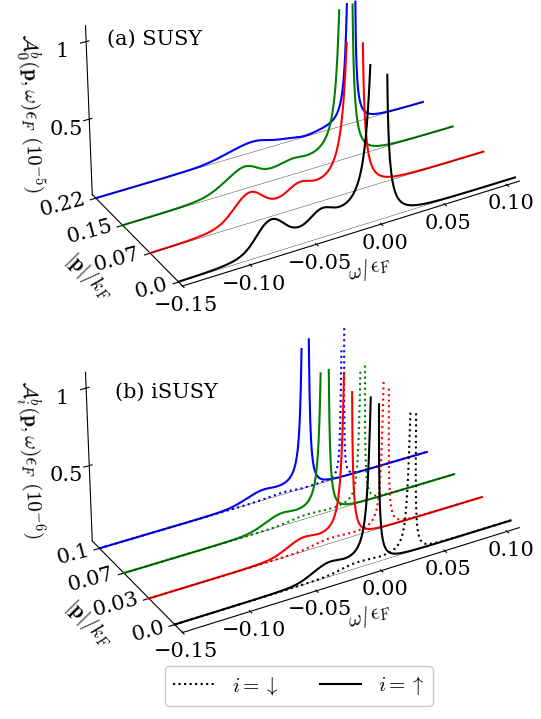}
\caption{\label{Fig: Spectral Function}
Plots of spectral functions
$\mathcal{A}^{b}_{i}(\bm{p},\omega)$
for all present species of bosons. The 
SUSY case is shown in (a), and the iSUSY case is illustrated in (b).}
\end{figure}

In Fig.~\ref{Fig: Spectral Function}, we show bosonic 
spectral functions $\mathcal{A}^{b}_{i}(\bm{p},\omega)$,
using parameters from Tab.~\ref{Tab:parameters_all}.
In panel (a), we depict the SUSY case, where there is only
one bosonic spectral function $\mathcal{A}^{b}_{0}(\bm{p},\omega)$, since the fermions carry the \textit{spin}. Three features appear, in order of increasing $\omega$, resulting from the $(0,\uparrow)$ Goldstino, the $(0,\downarrow)$ Goldstino, and the boson's ``independent'' particle energy (including mean-field shift).  
In panel (b), we show the iSUSY case, where the spectral function of each boson species $\mathcal{A}^{b}_{i}(\bm{p},\omega)$ is affected by a single Goldstino, either $(\uparrow, 0)$ or $(\downarrow, 0)$, since the bosons carry the \textit{spin}. Consequently, 
$\mathcal{A}^{b}_{i}(\bm{p},\omega)$ contains only two distinguishable features: one originating from the appropriate $(i,0)$ Goldstino, and the other from the bosons' ``independent'' particle energy. These features can be enhanced if $U_{ij}^{bf}$ is tuned experimentally.

Finally, when there are no Goldstinos present, either due to the system satisfying SUSY/iSUSY or the Goldstinos being completely submerged into the particle-hole continua, the spectral functions in Fig.~\ref{Fig: Spectral Function} retain only a single feature from the bosons' ``independent'' particle energies. The limit of $U_{ir}^{bf} \to 0$ is an example of this effect. 
Features attributable to the Goldstinos decrease as a function of $|\bm{p}|/k_F$. Due to the approximations made, the self-energy presented here is not valid beyond the momentum at which the Goldstino's feature would otherwise be enveloped by its own incoherent background.
These results can be leveraged for experimental detection of SUSY and iSUSY, including their distinguishing features, via RF~\cite{Stewart_2008,Koschorreck_2012,Jin_2015} or microwave-~\cite{Li_2024} based momentum-resolved spectroscopy. This applies to both the bosonic spectral functions, discussed here, and the fermionic ones analyzed in the End Matter.


\textit{Conclusions and Outlook\textemdash} We  
presented a theory where generalized supersymmetry (gSUSY) is broken and Goldstinos emerge from a Bose-Fermi mixture where component particles 
have internal \textit{spin} states leading to a maximum of four Goldstino flavors.
Further, we have differentiated gSUSY into two cases: supersymmetry (SUSY), in which fermions carry the \textit{spin}, and the inverse-supersymmetry (iSUSY), in which bosons carry the \textit{spin}. We then show that Goldstinos differ for SUSY an iSUSY cases in experimentally relevant ways. We established the regimes in which the Goldstinos are stable excitations with respect to the particle-hole continuum, and we identified features of their existence in radio-frequency momentum resolved spectroscopy via an analysis of the constituent bosonic and fermionic spectral functions. We showed that these features can be enhanced when Bose-Fermi interactions are experimentally tuned. We emphasized that simulating iSUSY is a very exotic feature of Bose-Fermi mixtures with \textit{spin} degrees of freedom, which is not found even in beyond standard model physics.
As an outlook, we point out that signatures of Goldstinos and anti-Goldstinos should also emerge in dynamical structure factor tensors, which can now be routinely measured.
 
\textit{Acknowledgments\textemdash} 
C.A.R.S.d.M. acknowledges support
from the Mercator Fellowship of the German Research Foundation (DFG) through the Collaborative Research Center SFB/TR185 (Project No. 277625399).

\bibliography{apssamp}

\appendix
\onecolumngrid
\section*{End Matter}
\twocolumngrid


\renewcommand{\theequation}{A\arabic{equation}}
\setcounter{equation}{0}

\textit{Supercharge density correlator \textemdash} 
We begin by defining the imaginary time correlator for 
the superchange density $q_{ij}(\bm{x}) = b^\dagger_i(\bm{x})f_j(\bm{x})$ as
\begin{equation}
    \mathcal{G}_{ijnm}(\bm{p},\tau)=-\left\langle\mathcal{T}_\tau q_{ij}(\bm{p},\tau)q^\dagger_{nm}(\bm{p},0)\right\rangle
\end{equation}
in which $\mathcal{T}_\tau$ is the imaginary time ordering operator.  
Using Wick's theorem, we can approximate this correlator by 
\begin{eqnarray}
    \nonumber
    \mathcal{G}_{ijnm}(\bm{p},\tau)&=&-\iint\frac{d^3\bm{k}d^3\bm{k}'}{(2\pi)^6}\left\langle\mathcal{T}_\tau b_i^\dagger(\bm{k},\tau)b_n(\bm{k}',0)\right\rangle \\* 
    &\times&\left\langle\mathcal{T}_\tau f_j(\bm{k}+\bm{p},\tau)f^\dagger_m(\bm{k}'+\bm{p},0)\right\rangle.
\end{eqnarray}
In the absence of \textit{spin}-flips in the Hamiltonian, this term is diagonal in the 
\textit{spin}-index of bosons and fermions. As such, we redefine the correlator's non-zero elements $\mathcal{G}_{ijij}(\bm{p},\tau)\equiv\mathcal{G}_{ij}(\bm{p},\tau)$ for ease of notation and show their structure within the RPA approximation in Fig. \ref{Fig: Goldstino Diagrams}.

\renewcommand{\theequation}{B\arabic{equation}}
\setcounter{equation}{0}

\textit{The Goldstino Pole and the Order of $\Lambda_{ij}$\textemdash}To decompose the denominator of Eq.~(\ref{Eq: RPA}) into $\bm{p}=\bm{0}$ and $\bm{p}\neq\bm{0}$ components, we first approximate the zero-momentum solution to first order in $|\chi|\ll1$. For $\bm{p}=\bm{0}$, this pole is located at the value of $\omega$ which satisfies
\begin{equation}
    0=\frac{1}{U^{bf}_{ij}}-\int\frac{d^3\bm{k}}{(2\pi)^3}\frac{n_{ij}(\bm{k},\bm{0})}{-\frac{\chi|\bf{k}|^2}{2m^*}+\bar{\omega}}
\end{equation}
where $\bar{\omega}=\omega+\xi^b_i(\bm{0})-\xi^f_j(\bm{0})$. Here, we consider only solutions yielding stable Goldstinos, which fall outside of the particle-hole continuum defined by
\begin{equation}
\label{Eq: add cont def}
\begin{cases}
\bar{\omega}\leq0, & \chi<0\\
\bar{\omega}=0, & \chi=0\\
\bar{\omega}\geq0, & \chi>0 .
\end{cases}
\end{equation}
The conditions above lead to those in Eq.~(\ref{Eq: Continuum definition}), when $\chi \ne 0$.

In the region where Goldstinos are stable, we expand the integrand in $\chi$, leading to 
\begin{eqnarray}
\nonumber
    0&=&\frac{1}{U^{bf}_{ij}}-\frac{N_{ij}}{\bar{\omega}}-\frac{\chi}{\bar{\omega}^2}\int\frac{d^3\bm{k}}{(2\pi)^3}\frac{|\bm{k}|^2}{2m^*}n_{ij}(\bm{k},\bm{0})\\*
    &-&\sum_{l=0}^\infty\frac{\chi^{l+2}}{\bar{\omega}^{l+3}}\int\frac{d^3\bm{k}}{(2\pi)^3}\left(\frac{|\bm{k}|^2}{2m^*}\right)^{l+2}n_{ij}(\bm{k},\bm{0}).
\end{eqnarray}
Here, we note that the integral in the first line is equivalent to $N_{ij}\langle\mathcal{E}_{ij}\rangle$, while those in the second line are higher-order moments thereof, all of which are $O(\chi^0)$. This results in 
\begin{equation}
0=\frac{1}{U^{bf}_{ij}N_{ij}}-\frac{1}{\bar{\omega}}-\frac{\chi\langle\mathcal{E}_{ij}\rangle}{\bar{\omega}^2}+O(\chi^2),
\end{equation}
which has the solution $\bar{\omega}=U^{bf}_{ij}N_{ij}+\chi\langle\mathcal{E}_{ij}\rangle$ up to first order in $\chi$, which suggests that $\omega=\Omega_{ij}$ as defined in Eq.~(\ref{Eq: Omega}).

Motivated by the zero-momentum solution, we shift our focus to the arbitrary momentum case. In rewriting $G^\text{MF}_{ij}$ for $\chi\neq0$ as
\begin{equation}
    G^{MF}_{ij}(\bm{p},\omega)=\int\frac{d^3\bm{k}}{(2\pi)^3}\frac{n_{ij}(\bm{k},\bm{p})}{\omega-\zeta_{ij}(\bm{k},\bm{p})+\text{i}0^+},
\end{equation}
we obtain
\begin{equation}
    1-U^{bf}_{ij}G^{MF}_{ij}(\bm{p},\omega)=\frac{1}{U^{bf}_{ij}N_{ij}}\left[\omega-\Omega_{ij}-\Lambda_{ij}(\bm{p},\omega)\right]
\end{equation}
where the non-zero momentum contribution $\Lambda_{ij}(\bm{p},\omega) $ is defined in Eq.~(\ref{Eq: Lambda Def}).
Finally, to verify its agreement with our zero-momentum solution, we must show that $\Lambda_{ij}(\bm{0},\Omega_{ij})$ is at least of order $\chi^2$. Evaluating $\Lambda_{ij}(\bm{0},\Omega_{ij})$ outside of the continuum and
expanding it to the lowest order of $\chi$ results in the first non-zero term being
\begin{equation}
    \Lambda_{ij}(\bm{0},\Omega_{ij})\approx\frac{\chi^2}{U^{bf}_{ij}N_{ij}^2}\int\frac{d^3\bm{k}}{(2\pi)^3}n_{ij}(\vec{k},\bm{0})\left(\langle\mathcal{E}_{ij}\rangle-\frac{|\vec{k}|^2}{2m^*}\right)^2.
\end{equation}
Here, it is clear that $\Lambda_{ij}(\bm{0},\Omega_{ij})=O(\chi^2)$ and is thus negligible compared to $\Omega_{ij}=O(\chi)$.


\renewcommand{\theequation}{C\arabic{equation}}
\setcounter{equation}{0}

\textit{The Residue of the Pole\textemdash}
We focus on the pole of the RPA correlator given in Eq.~(\ref{Eq: RPA}), which corresponds to the $(i,j)$ Goldstino's propagator
\begin{equation}
G^q_{ij}(\bm{p},\omega)=\frac{Z_{ij}(\bm{p})}{\omega-\mathcal{D}_{ij}(\bm{p})+\text{i}0^+}\approx\frac{U^{bf}_{ij}N_{ij} G^\text{MF}_{ij}(\bm{p},\omega)}{\omega-\Omega_{ij}-\Lambda_{ij}(\bm{p},\omega)}
\label{Eq: Gq compare}
\end{equation}
defined in Eq.~(\ref{Eq: Gq}), with $\Lambda_{ij}(\bm{p},\omega)$ defined in Eq.~(\ref{Eq: Lambda Def}).
From the expression above, we extract the Goldstino dispersion 
$\mathcal{D}_{ij}(\bm{p})$ as the integer zero of the denominator at 
$\omega = \mathcal{D}_{ij}(\bm{p})$, that is, $ \mathcal{D}_{ij}(\bm{p})- \Omega_{ij} - 
\Lambda_{ij} (\bm{p},\mathcal{D}_{ij}(\bm{p})) = 0$, and the residue of the pole (akin to the wavefunction renormalization) is
\begin{equation}
    Z_{ij}(\bm{p})=\frac{U^{bf}_{ij}N_{ij} G^\text{MF}_{ij}(\bm{p},\mathcal{D}_{ij}(\bm{p}))}{1-\left.\partial\Lambda_{ij}(\bm{p},\omega)/\partial\omega\right|_{\omega=\mathcal{D}_{ij}(\bm{p})}}
\end{equation}
after the application of L'Hôpital's rule. This can be algebraically simplified to yield Eq.~(\ref{Eq: residue}). 

Here, we discuss two limiting cases of the residue $Z_{ij}(\bm{p})$. First, in the zero-momentum limit in which $\bm{p}\rightarrow\bm{0}$ and $\mathcal{D}_{ij}(\bm{0})\rightarrow\Omega_{ij}$, we find that $Z_{ij}(\bm{0})\approx U^{bf}_{ij}N_{ij} G^\text{MF}_{ij}(\bm{0},\Omega_{ij})\left[1+O(\chi^2)\right]$, in agreement with the expected result from Eq.~(\ref{Eq: Gq compare}).
Second, in the limit where the Goldstino dispersion $\mathcal{D}_{ij}(\bm{p})$ approaches the particle-hole continuum at $\bm{p} \rightarrow \bm{p}^*$,
that is, $\mathcal{D}_{ij}(\bm{p}^*)=\zeta_{ij}(\bm{0},\bm{p}^*)$, the residue vanishes as
\begin{equation}
    \lim_{\bm{p}\rightarrow\bm{p}^*}Z_{ij}(\bm{p})=\frac{1}{U^{bf}_{ij}}\lim_{\bm{p}\rightarrow\bm{p}^*}\left[\mathcal{D}_{ij}(\bm{p}^*)-\zeta_{ij}(\bm{0},\bm{p}^*)\right]=0.
\end{equation}
This indicates that when stable Goldstinos approach their
particle-hole continua, their residue approaches zero and they contribute less to the constituent particle spectral functions.


\renewcommand{\theequation}{D\arabic{equation}}
\setcounter{equation}{0}

\textit{Converting $U^{bf}_{rs}$ into $a^{bf}_{rs}$\textemdash} The Bose-Fermi interactions $U^{bf}_{rs}$ can 
be converted into the Bose-Fermi scattering lengths 
$a^{bf}_{rs}$ via the two-body Lippmann–Schwinger equation
\begin{equation}
    \label{Eq: LS equation}
    U^{bf}_{rs}=\frac{2\pi a^{bf}_{rs}}{m^*}\left(1-\frac{2}{\pi }k_c a^{bf}_{rs}\right)^{-1},
\end{equation}
wherein $a^{bf}_{rs}$ is the s-wave scattering length between the relevant internal states and $k_c \sim 1/R_{bf}$ is the momentum cutoff, reflecting the range $R_{bf}$ of the interactions.  
For a typical Bose-Fermi particle separation $\ell_{BF}$, 
the ratio between the interaction range volume 
$4 \pi R_{bf}^3/3$ and the typical Bose-Fermi interparticle separation volume  $4 \pi \ell_{bf}^3/3$ must be $(R_{bf}/\ell_{bf})^3 \ll 1$. We choose the value
of $R_{bf}/\ell_{bf} =1/10$, leading to $k_c a_{rs}^{bf} \ll 1$.


\renewcommand{\theequation}{E\arabic{equation}}
\setcounter{equation}{0}

\begin{figure}
\includegraphics[width=8.5 cm]{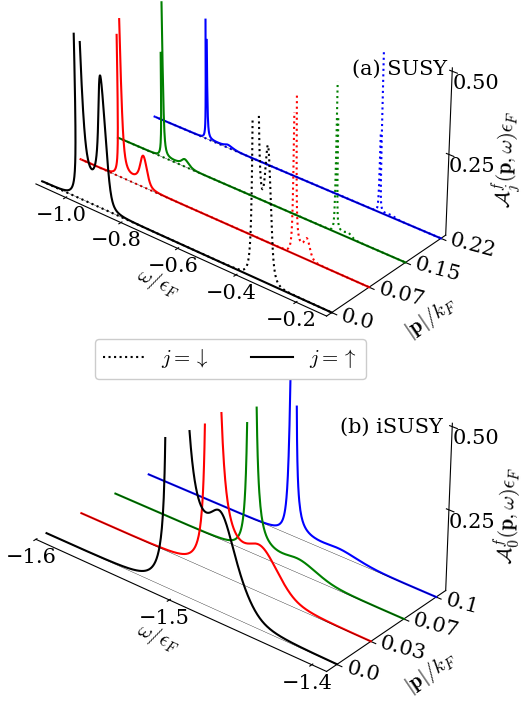}
\caption{\label{Fig: Fermion Spectral Function} Plots of spectral functions $\mathcal{A}^{f}_{j}(\bm{p},\omega)$ for all present species of fermions. The SUSY case is shown in (a), and the iSUSY case is illustrated in (b).}
\end{figure}

\textit{The Fermion Spectral Function\textemdash}
While we have already discussed the resulting spectral functions for bosons in the main text, here we perform the same analysis for fermions. The spectral function for fermions in state $j$ is $\mathcal{A}^{f}_{j}(\bm{p},\omega)=-\Im \left[ G_{j}^{f}(\bm{p},\omega)\right]/\pi$, where $G^{f}_{j}(\bm{p},\omega)=[\omega-\xi^{f}_{j}(\bm{p})-\sum_r\Sigma^f_{rj}(\bm{p},\omega)+\text{i}0^+]^{-1}$  is the fermion's retarded propagator. Here,
\begin{equation}
\Sigma^f_{rj}(\bm{p},\omega)
= 
- \left(U^{bf}_{rj}\right)^2
\int \frac{d^3\bm{k}}{(2\pi)^3}
Z_{rj}(\bm{k})\\ 
\Pi_{f} (\bm{k}, \bm{p}, \omega),
\end{equation}
where the boson-Goldstino term 
\begin{equation}
\Pi_{f} (\bm{k}, \bm{p}, \omega) =
\frac{n_f\left(\mathcal{D}_{rj}\left(\bm{k}\right)\right)+n_b\left(-\xi^b_r\left(\bm{p}-\bm{k}\right)\right)}{\omega-\xi^b_r\left(\bm{p}-\bm{k}\right)-\mathcal{D}_{rj}\left(\bm{k}\right)+\text{i}0^+}
\end{equation}
accounts for the fermion's self-energy corrections presented in Fig. \ref{Fig: Self Energy Diagram}. Utilizing the parameters listed in Table~\ref {Tab:parameters_all}, we show the numerically evaluated spectral function for fermions in Fig. \ref{Fig: Fermion Spectral Function}, in which we also see manifestations of the Goldstinos similar to those found in the spectral function of bosons discussed in the main text. In the SUSY case, there are two species of fermions, each of which has its own spectral function with two features: one originating from the relevant $(0,j)$ Goldstino and the other from the ``independent'' fermion particle energy. Using the same parameters as we did for bosons, we find that the iSUSY spectral function for fermions only shows two distinguishable features. Here, the $(\uparrow,0)$ and $(\downarrow,0)$ Goldstino features overlap, becoming unresolvable. Such an effect is not a necessary result of the theory, it is rather 
a consequence of the choice of parameters.

\end{document}